\documentclass[reprint,prl,superscriptaddress]{revtex4-1}

\usepackage{amsmath}
\usepackage{amsfonts}
\usepackage{amssymb}
\usepackage{graphicx}
\usepackage{moreverb}
\usepackage{url}
\usepackage{epsf}
\usepackage{color}
\usepackage{multirow}
\usepackage{subfigure}

\begin{document}

\title{Enhanced configurational entropy in high-density nanoconfined bilayer ice}
\author{Fabiano~Corsetti}
\email[E-mail: ]{fabiano.corsetti08@imperial.ac.uk}
\affiliation{CIC nanoGUNE, 20018 Donostia-San Sebasti\'{a}n, Spain}
\affiliation{Department of Materials and the Thomas Young Centre for Theory and Simulation of Materials, Imperial College London, London SW7 2AZ, United Kingdom}
\author{Jon~Zubeltzu}
\affiliation{CIC nanoGUNE, 20018 Donostia-San Sebasti\'{a}n, Spain}
\author{Emilio~Artacho}
\affiliation{CIC nanoGUNE, 20018 Donostia-San Sebasti\'{a}n, Spain}
\affiliation{Theory of Condensed Matter, Cavendish Laboratory, University of Cambridge, Cambridge CB3 0HE, United Kingdom}
\affiliation{Basque Foundation for Science Ikerbasque, 48011 Bilbao, Spain}
\affiliation{Donostia International Physics Center, 20018 Donostia-San Sebasti\'{a}n, Spain}
\date{\today}

\begin{abstract}
A novel kind of crystal order in high-density nanoconfined bilayer ice is proposed from molecular dynamics and density-functional theory simulations. A first-order transition is observed between a low-temperature proton-ordered solid and a high-temperature proton-disordered solid. The latter is shown to possess crystalline order for the oxygen positions, arranged on a close-packed triangular lattice with $AA$ stacking. Uniquely amongst the ice phases, the triangular bilayer is characterized by two levels of disorder (for the bonding network and for the protons) which results in a configurational entropy twice that of bulk ice.
\end{abstract}

\maketitle

The exceptional polymorphism of water in its crystal state (with 16 known bulk phases~\cite{0036-021X-75-1-R04, IceXIIIandXIV, IceXV, IceXVI}) arises from two properties of its molecular bonding: Firstly, the tendency to form an open and flexible network, which allows for many different bonding topologies. Secondly, the possibility of proton disorder, which results in a distinction between a high-temperature disordered phase and a low-temperature ordered phase on the same bonding network. Such a transition has been observed for almost all phases (Ih-XI, III-IX, V-XIII~\cite{IceXIIIandXIV}, VI-XV~\cite{IceXV}, VII-VIII, XII-XIV~\cite{IceXIIIandXIV}). The proton disorder introduces a configurational entropy following the Bernal-Fowler ice rules~\cite{Bernal1933}. Pauling's simple estimation~\cite{Pauling1935} of the entropy, $W = \left ( 3/2 \right)^N$ (where $W$ is the number of microstates for $N$ water molecules), is surprisingly accurate; indeed, recent numerical calculations~\cite{Herrero2014} show only a very small variation between all proton-disordered bulk phases from $W \simeq 1.504^N$ (XII) to $W \simeq 1.524^N$ (VI).

Nanoconfined 2D water, of interest in fields ranging from biochemistry to nanotechnology, has shown a similar capacity for polymorphism (see review and references in Zhao {\em et al.}~\cite{Zhao2014a} and Corsetti {\em et al.}~\cite{monolayer}). When considered, the configurational entropy of nanoconfined ices has been found to be either non-extensive, leading to an intrinsically ordered phase (e.g., monolayer square ice~\cite{Zangi2003a, Koga2005, monolayer}), or less than or equal to that of the disordered bulk phases (e.g., monolayer~\cite{monolayer} and bilayer~\cite{Koga1997} honeycomb ice). No order--disorder transition on the same network has yet been identified.

In this paper, we investigate the phases of bilayer ice obtained by confining water in one dimension between infinite plates with sub-nm separation. Different properties of the liquid and solid bilayer have been explored in previous experimental~\cite{Algara-Siller2015} and computational studies~\cite{Koga1997, Meyer1999, Koga2000, Zangi2003, Zangi2003a, Bai2003, Kumar2005, Koga2005, Giovambattista2006, Kumar2007, Giovambattista2009, Johnston2010, Han2010, Bai2012, Mosaddeghi2012, Ferguson2012, Algara-Siller2015}, all of which have made use of empirical force-field models of water (mW~\cite{Johnston2010}, SPC/E~\cite{Giovambattista2006, Giovambattista2009, Mosaddeghi2012, Ferguson2012, Algara-Siller2015}, TIP4P~\cite{Koga1997, Meyer1999, Koga2000, Koga2005}, TIP4P/Ice~\cite{Johnston2010}, TIP5P~\cite{Zangi2003, Zangi2003a, Bai2003, Kumar2005, Kumar2007, Han2010, Bai2012}, ST2~\cite{Bai2003}). The majority of studies report a low-density honeycomb bilayer crystal~\cite{Koga1997, Koga2000, Bai2003, Koga2005, Giovambattista2006, Giovambattista2009, Johnston2010, Han2010, Bai2012, Ferguson2012}. Han {\em et al.}~\cite{Han2010} distinguish a separate high-density solid (also seen by Zangi and Mark~\cite{Zangi2003}), which they define as rhombic and suggest could be a hexatic phase. Bai and Zeng~\cite{Bai2012} describe a similar solid, but define it instead as very-high-density bilayer amorphous ice.

\begin{figure}
\includegraphics[width=0.49\textwidth]{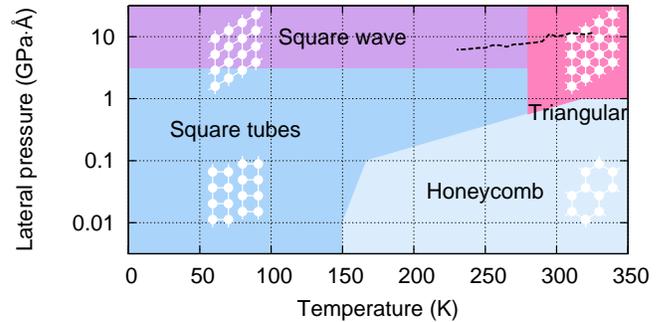}
\caption{Schematic diagram of the stable crystalline phases of bilayer ice obtained in this study for a confinement width of 8~\AA. The diagram is semi-quantitative, combining enthalpy calculations by DFT-AIRSS with the high-temperature phase transition to triangular ice obtained by MD (pressures along the heating path are given by the thick dashed black line). See Ref.~\cite{bilayer-liquid} for a more detailed phase diagram of the high-temperature range from MD, including the melting line which is omitted here for simplicity.}
\label{fig:pd}
\end{figure}

We look in detail at this high-density phase, showing that the oxygens form a fixed and ordered triangular lattice which only becomes apparent when averaging over a sufficiently long timescale. The high degree of proton disorder allowed by the lattice gives the bilayer triangular phase several interesting and unique properties; in particular, it possesses twice the entropy of the bulk phases without breaking the Bernal-Fowler ice rules. We also demonstrate the transition to at least one ordered phase at low temperature, with a possible distinct ordered phase appearing at very high lateral pressure, as shown in Fig.~\ref{fig:pd}.

We use two models at different levels of theory for our calculations. Firstly, we perform classical molecular dynamics (MD) simulations of 294 water molecules with the TIP4P/2005~\cite{Abascal2005} force-field model in the LAMMPS~\cite{Plimpton1995} code. We use a time step of 0.5~fs, a cutoff of 12~\AA\ for Lennard-Jones (L-J) interactions, and the particle--particle particle--mesh~\cite{pppm} (PPPM) method for long-range Coulombic interactions. Secondly, we perform an {\em ab initio} random structure search~\cite{Pickard2011} (AIRSS) procedure to identify the lowest-enthalpy configurations for small unit cells (we test cells of four and eight molecules). The AIRSS uses density-functional theory (DFT) calculations in the SIESTA~\cite{Ordejon1996,Soler2002} code with a fully non-local exchange and correlation functional that gives an accurate description of van der Waals interactions in water from first principles~\cite{Corsetti2013b}. A detailed description of the DFT and AIRSS methodologies can be found in our previous study of monolayer ice~\cite{monolayer}. For both the MD and DFT simulations the confining walls are described by the same classical L-J 9-3 potential~\cite{monolayer}, and the simulation cell is periodic in all directions with a buffer region in $z$ of 11~\AA.

\begin{figure}
\includegraphics[width=0.49\textwidth]{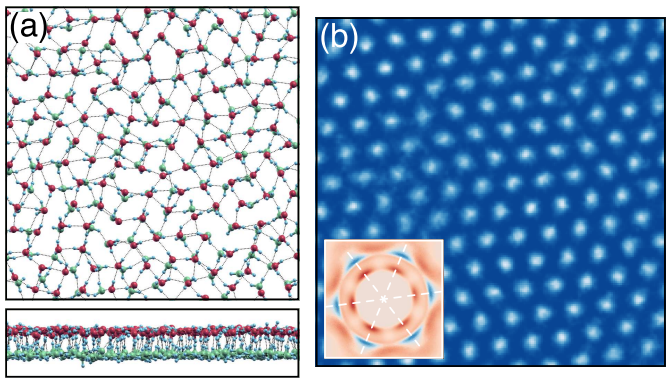}
\caption{The structure of triangular bilayer ice from MD at 310~K. (a) Instantaneous snapshot, shown parallel (top panel) and perpendicular (bottom panel) to the confinement direction. The cell side is 34.40~\AA. The oxygens in the two layers are colored differently to help visualization. (b) Oxygen positions averaged over the entire run; both layers are included. The inset shows the average nearest-neighbor positions with respect to a central molecule; in this case both oxygens (in blue) and protons (in red) are shown. The cell side for the inset is 7~\AA.}
\label{fig:triangular}
\end{figure}

Fig.~\ref{fig:triangular} shows a representative example of the triangular phase obtained by MD. The simulation is carried out in a fixed square cell with an area per molecule of 4.02~\AA$^2$ and a confinement width of 8~\AA. The cell is first equilibrated with a Berendsen thermostat for 60~ns, after which statistics are collected for 100~ps in the $NVE$ ensemble.

Any instantaneous snapshot of the system (Fig.~\ref{fig:triangular}(a)) seems to confirm the definition of an amorphous solid. Although the particles are densely packed, they are not arranged on a clear lattice and there is no visible order in the hydrogen-bonding network; instead, there is an irregular combination of squares, pentagons, hexagons, and so on (similarly to low-density amorphous bilayer ice~\cite{Koga2000, Bai2003, Koga2005, Bai2012}). If quenched to low temperature the system would therefore freeze into this amorphous arrangement. However, the bonding network constantly rearranges during the course of the simulation, and the molecules are pulled towards different neighbors depending on the instantaneous bonding arrangement. When averaging the oxygen positions over the entire 100~ps (Fig.~\ref{fig:triangular}(b)) a very clear triangular lattice emerges. The unit cell is found to be hexagonal to a good approximation (with errors of 1\% in the unit cell angle and 0.3\% in the $a/b$ ratio, which can be attributed to finite size effects due to the fixed supercell). The inset of the figure shows that every molecule is at the centre of a hexagonal cage with six in-plane neighbors; the hydrogen bonds (shown by the inner circle of protons bonding to the central molecule) fluctuate between all six neighbors with no preferential direction. It is also important to note that the bilayer has an $AA$ stacking. Each molecule therefore maintains a constant bond to the other layer, and three fluctuating in-plane bonds. The number of molecules with defective configurations is reasonably small, $<$5\% of the total.

\begin{figure}
\includegraphics[width=0.49\textwidth]{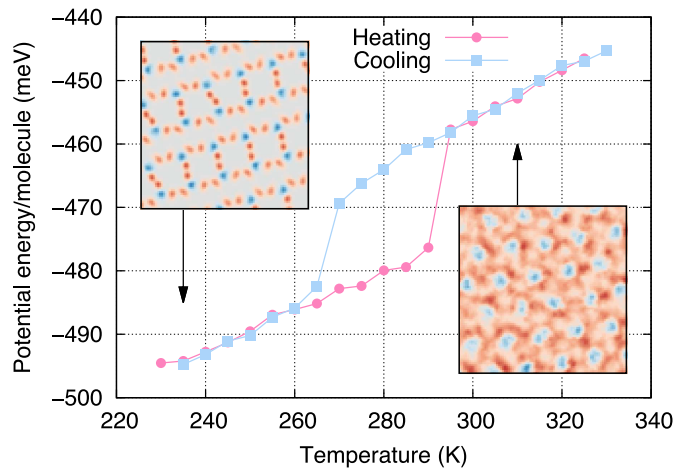}
\caption{Potential energy per molecule as a function of temperature from MD. The cell shape and size and the number of molecules is kept constant. The insets show the structure of square tubes bilayer ice at 235~K and triangular bilayer ice at 310~K. The position of the oxygens (in blue) and protons (in red) is averaged over the entire run for both phases, with both layers being included. A $\left ( 15 \times 15 \right)$~\AA$^2$ portion of the entire cell is shown.}
\label{fig:loop_inset}
\end{figure}

The molecular bonding arrangement of bilayer triangular ice is consistent with the expected sixfold orientational order of a hexatic phase; we discuss this in more detail elsewhere~\cite{bilayer-liquid}. It is interesting to note that a hexatic ice monolayer has previously been predicted and analyzed using a coarse-grained model~\cite{monolayer-hexatic}; however, in that case the number of hydrogen bonds was found to be greatly reduced, and, therefore, is significantly different to what we observe.

The fluctuating hydrogen bonds suggest a large configurational entropy and the possibility of an order--disorder transition upon cooling. We first verify this empirically within our MD simulation, by starting from a high-temperature configuration (prepared as described previously) and cooling in steps of 5~K. Each step is carried out by re-equilibrating the system with a Berendsen thermostat for 5~ns, and then collecting $NVE$ statistics for 100~ps. The reverse heating process is also performed independently. The two processes are shown in Fig.~\ref{fig:loop_inset}.

The sharp discontinuity and hysteresis loop in the potential energy of the system are indications of a first-order phase transition occurring at $T_c \simeq 280 \pm 15$~K. Below this temperature the triangular bilayer transforms into a phase characterized by rows of square-based tubes, with no hydrogen bonds between them~\cite{Bai2012}. The unit cell symmetry is lowered from hexagonal to rhombic (centred rectangular), with an angle of $\sim$106$^{\circ}$. The position of the protons as well as the oxygens in the square tubes is fixed for the entire simulation, denoting a proton-ordered phase with no configurational entropy. We can therefore estimate the configurational entropy of the triangular phase by equating it with the gain in internal energy at the transition (assuming a similar vibrational contribution). Since we are in the $NVE$ ensemble, $\Delta S = \Delta U/T_c \simeq 0.7 \pm 0.1$~$k_\mathrm{B}$/molecule. The configurational entropy is therefore almost twice the value for bulk ice, $S/N \simeq 0.4$~$k_\mathrm{B}$. Additional calculations in the $NP_{xy}T$ ensemble give a perfect qualitative and quantitative agreement both for $T_c$ and $\Delta S$ (estimated in this case as $\Delta H/T_c$); details are in the Supplemental Material.

\begin{figure}
\includegraphics[width=0.49\textwidth]{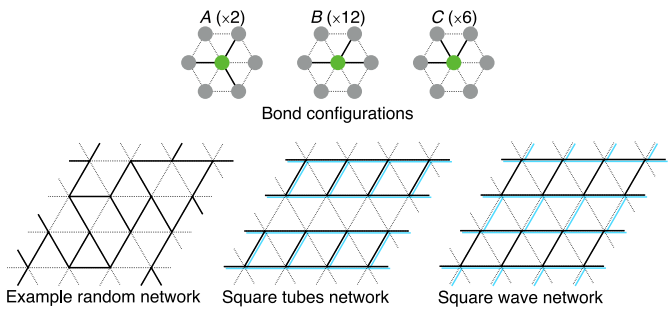}
\caption{All non-equivalent hydrogen-bond configurations around a site of the triangular lattice (symmetry degeneracy given in brackets). Some examples of extended networks are shown below; for the square tubes and square wave networks the bonds in the second layer are shown in light blue.}
\label{fig:bond_configs}
\end{figure}

The large entropy value can be understood by considering the generalization of the ice rules to the case of the triangular bilayer lattice. There are two levels of configurational entropy, in the arrangement of hydrogen bonds on the lattice and in the placement of protons along those bonds. For all other solid phases only the latter applies, since the bonding network is fixed.

Fig.~\ref{fig:bond_configs} shows the possible configurations of the three in-plane hydrogen bonds around a molecule. As already mentioned, the fourth bond is fixed and connects the molecule to the other layer. The bonding pattern in the two layers is therefore decoupled. From the random network example shown in the figure it can be seen how polygons of varying size and shape are created, resulting in the characteristic amorphous appearance of the instantaneous snapshots.

\begin{figure}
\includegraphics[width=0.49\textwidth]{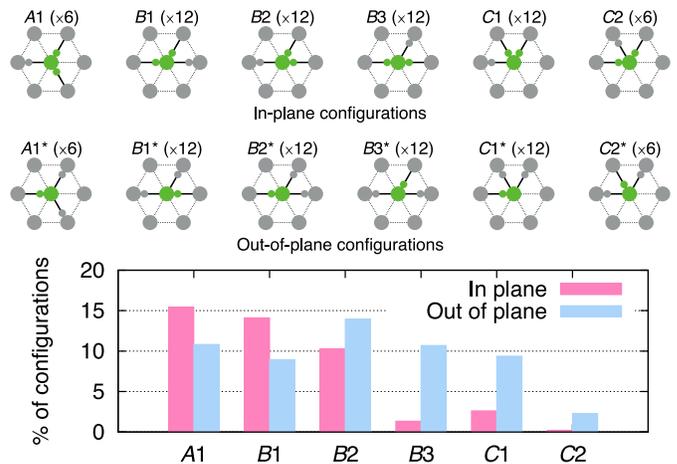}
\caption{All allowed non-equivalent proton configurations for a molecule on the triangular bilayer lattice (symmetry degeneracy given in brackets). Out-of-plane molecules donate a proton to the interlayer bond (not pictured). The frequency for each configuration, shown below, is calculated from MD simulations of the triangular phase.}
\label{fig:proton_configs}
\end{figure}

It should also be expected that low-energy configurations will result from particular regular bonding patterns; indeed, the square tubes network can be created on the triangular lattice entirely from $B$ configurations (see Fig.~\ref{fig:bond_configs}). This is an unusual pattern, made up of entirely disconnected networks with 1D periodicity. The regular pattern allows for a macroscopic distortion of the lattice (as opposed to the localized distortions of the molecules in the random network), which is enhanced by the weak interaction between tubes. This results in the change of symmetry observed in the simulation.

The second level of configurational entropy is the arrangement of protons on the bonding network (Fig.~\ref{fig:proton_configs}), such that each molecule has two short and two long OH distances. The two layers are now no longer independent, because the total number of in-plane and out-of-plane molecules in the system has to be equal, but they do not have to be equally distributed between the two layers. There are 12 non-symmetrically equivalent proton configurations for an individual molecule, and 120 distinguishable configurations in total.

It should be expected that not all proton configurations are energetically equivalent, and some might be effectively prohibited for this reason, thus reducing $S$. Table~SI in the Supplemental Material gives our Pauling-like estimate for the configurational entropy of the system depending on which proton configurations are allowed; we use a Monte Carlo method to calculate this, also detailed in the Supplemental Material. Including all possible configurations gives an unreasonably high value of $1.35 \pm 0.01$~$k_\mathrm{B}$. Indeed, if we analyze our MD simulations, as shown in the histogram in Fig.~\ref{fig:proton_configs}, we find that four configurations are almost entirely absent ($B$3, $C$1, $C$2, $C$2*). Interestingly, there is also a noticeable asymmetry between in-plane and out-of-plane configurations, with the latter being more permissive ($B$3* and $C$1* are allowed). Taking these restrictions into account, the Pauling estimate is $0.8 \pm 0.1$~$k_\mathrm{B}$, in good agreement with the value obtained previously from MD.

\begin{figure}
\includegraphics[width=0.49\textwidth]{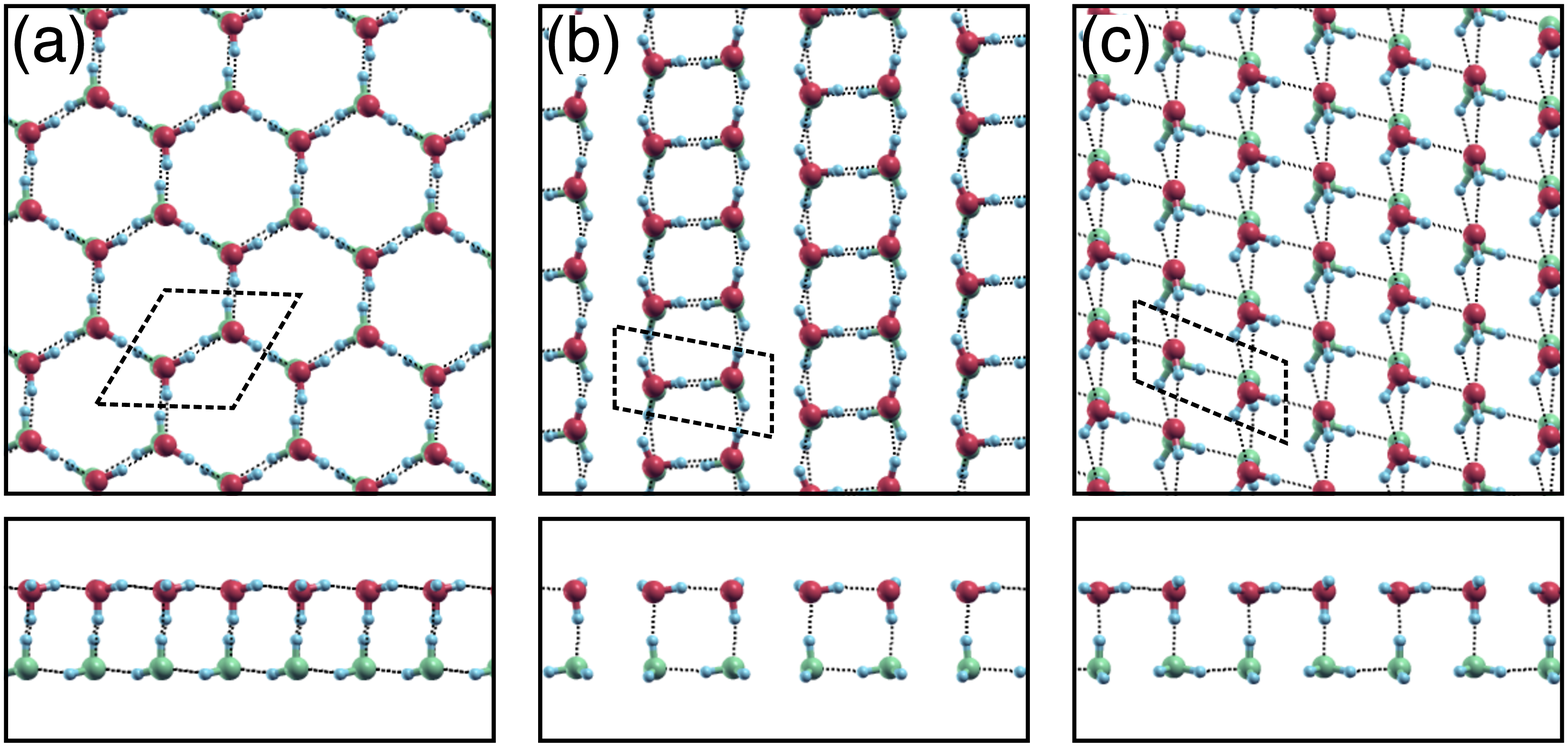}
\caption{Relaxed crystal structures for stable bilayer ice phases calculated by DFT and found with an AIRSS methodology. (a) Honeycomb, lateral pressure of 0.01~GPa$\cdot$\AA, area per molecule of 4.98~\AA$^2$. (b) Square tubes, lateral pressure of 0.01~GPa$\cdot$\AA, area per molecule of 4.03~\AA$^2$. (c) Square wave, lateral pressure of 10~GPa$\cdot$\AA, area per molecule of 3.83~\AA$^2$. The confinement width is 8~\AA. The four-molecule unit cell is shown.}
\label{fig:crystal}
\end{figure}

We conclude our analysis by returning to the overall phase diagram presented in Fig.~\ref{fig:pd}. Here we make use of first principles calculations with DFT and the AIRSS methodology to identify the lowest-enthalpy structures at different lateral pressures. Temperature dependence is obtained by estimating the phases' configurational entropy, and, hence, Gibbs free energy. Our previous study of monolayer ice has shown the contribution to free energy differences from vibrational effects to be small~\cite{monolayer}.

The AIRSS recovers the low-density honeycomb bilayer crystal (Fig.~\ref{fig:crystal}(a)). Contrary to previous reports~\cite{Koga1997}, there is no noticeable distortion in the hexagonal unit cell, and so no preferential proton orientation. The derivation of the configurational entropy is therefore equivalent to that of Pauling for bulk ice (more than twice the previous estimate for the bilayer~\cite{Koga1997}).

Two high-density bilayer crystals, close in enthalpy, are predicted: square tubes ice (Fig.~\ref{fig:crystal}(b)), and a novel structure not found by MD which we refer to as square wave ice (Fig.~\ref{fig:crystal}(c)). The latter has a similar bonding network to the tubes, but with the two layers shifted relative to each other; the whole crystal is now connected into a single network, and when viewed perpendicular to the confinement direction gives a square wave pattern. This is illustrated schematically in Fig.~\ref{fig:bond_configs}. The rhombic distortion with respect to the hexagonal cell is less pronounced for the wave than for the tubes, with unit cell angles of 112$^{\circ}$ and 101$^{\circ}$, respectively.

The square wave and the square tubes phases are predicted to be ordered on both levels. This is because the bonding network is fixed, and all the examples recovered from the AIRSS make use of only four proton configurations ($B$1, $B$1*, $B$2, $B$2*). Two other possible configurations compatible with the network are not found ($B$3, $B$3*). It can easily be shown that this results in a non-extensive entropy. It is important to note that the specific configuration of protons in the square tubes obtained by MD using TIP4P/2005 is in agreement with the lowest-energy configuration found by DFT. A previous study using TIP5P instead gives a different configuration~\cite{Bai2012}. This is a similar result to what was found for the square monolayer~\cite{monolayer}, and a further indication of the accuracy of the TIP4P/2005 force-field model.

Recent experimental observations of ice confined between graphene sheets~\cite{Algara-Siller2015} show a square configuration for the monolayer up to the trilayer. While DFT and some force-field models are in excellent agreement with experiment for the monolayer~\cite{monolayer}, we have been unsuccessful in obtaining the square bilayer; calculations using both DFT and TIP4P/2005 find it to be unstable with respect to the high-density phases discussed here. This suggests interesting effects originating either from deeper levels of theory, or considerations such as the confinement and the dynamics of formation, meriting further study both from experiment and simulation~\cite{bilayer-new}.

\begin{acknowledgments}
This work was partly funded by grants FIS2012-37549-C05 from the Spanish Ministry of Science, and Exp.\ 97/14 (Wet Nanoscopy) from the Programa Red Guipuzcoana de Ciencia, Tecnolog\'{i}a e Innovaci\'{o}n, Diputaci\'{o}n Foral de Gipuzkoa. We thank Jos\'{e} M. Soler and M.-V. Fern\'{a}ndez-Serra for useful discussions. The calculations were performed on the arina HPC cluster (Universidad del Pa\'{i}s Vasco/Euskal Herriko Unibertsitatea, Spain). SGIker (UPV/EHU, MICINN, GV/EJ, ERDF and ESF) support is gratefully acknowledged.
\end{acknowledgments}

\end{document}